\PassOptionsToPackage{pdfpagelabels=false}{hyperref}
\RequirePackage{rotating}
\documentclass[usenatbib]{mnras}
\usepackage{xcolor, graphicx}
\usepackage{amsmath}
\usepackage{amssymb}
\usepackage{soul}
\usepackage{txfonts}
\usepackage{journal_names}
\usepackage{caption, booktabs}
\usepackage{todonotes}
\usepackage[export]{adjustbox}
\usepackage{longtable}
\usepackage{rotating}
\usepackage{lscape}

\newcommand{\Msun}{\rm M_\odot}
\newcommand{\kms}{\mbox{km s$^{-1}$}}
\newcommand{\vlos}{V_{\rm los}}
\newcommand{\SBunit}{\mbox{mag arcsec$^{-2}$}}
\newcommand{\re}{R_{\rm e}}

\newcommand{\unsim}{\mathord{\sim}}

\def\arcmin{\hbox{$^\prime$}}
\def\arcsec{\hbox{$^{\prime\prime}$}}

\def\micron{{{$\mu$}m}}
\def\Sersic/{{S\'ersic}}

\title[Shells in NGC~474 -- KCWI]
{NGC~474 as viewed with KCWI: diagnosing a shell galaxy}
\author[Alabi~et~al.~ ]
{Adebusola B. Alabi$^{1,2}$\thanks{email: aalabi@ucsc.edu}, Anna Ferr{\'e}-Mateu$^{3}$, Duncan A. Forbes$^{2}$, Aaron J. Romanowsky$^{1,4}$,
\newauthor  Jean P. Brodie$^{1,2}$\\
\\
$^{1}$ University of California Observatories, 1156 High Street, Santa Cruz, CA 95064, USA\\
$^{2}$ Centre for Astrophysics \& Supercomputing, Swinburne University, Hawthorn VIC 3122, Australia\\
$^{3}$ Institut de Ciencies del Cosmos (ICCUB), Universitat de Barcelona (IEEC-UB), E02028 Barcelona, Spain\\
$^{4}$ Department of Physics and Astronomy, One Washington Square, San Jos\'e State University, San Jose, CA 95192, USA\\
}

\begin{document}

\date{Accepted today}
\pagerange{\pageref{firstpage}--\pageref{lastpage}} \pubyear{2020}
\maketitle

\label{firstpage}
\begin{abstract}
We present new spectra obtained using Keck/KCWI and perform kinematics and stellar population analyses of the shell galaxy NGC~474, from both the galaxy centre and a region from the outer shell. We show that both regions have similarly extended star formation histories although with different stellar population properties. The central region of NGC~474 is dominated by intermediate-aged stars ($8.3\pm0.3$~Gyr) with subsolar metallicity ([$Z$/H]$=-0.24\pm0.07$~dex) while the observed shell region, which hosts a substantial population of younger stars, has a mean luminosity-weighted age of $4.0\pm0.5$~Gyr with solar metallicities ([$Z$/H]=$-0.03\pm0.09$~dex). Our results are consistent with a scenario in which NGC~474 experienced a major to intermediate merger with a log$(M_*/\Msun)\unsim10$ mass satellite galaxy at least $\unsim2$~Gyr ago which produced its shell system. This work shows that the direct spectroscopic study of low-surface brightness stellar features, such as shells, is now feasible and opens up a new window to understanding galaxy formation and evolution.
\end{abstract}


\section{Introduction}
NGC~474 is perhaps best known for its spectacular system of stellar shells \citep{Malin_1983, Turnbull_1999, Forbes_1999, Duc_2015}. Shells, alongside similar faint stellar tidal features, e.g. streams, loops, plumes, rings, etc. are telltale signatures of the hierarchical assembly of present-day galaxies and are believed to have formed during galaxy mergers or close galaxy interactions \citep{Schweizer_1980, Quinn_1984, Dupraz_1986, Hernquist_1988, Thomson_1990, Thomson_1991, Gonzalez_2005, Pop_2018, Kado_2018, Karademir_2019}. 

NGC~474, classified as a type II shell system \citep{Prieur_1988}, is the dominant S0 galaxy \citep{Pierfederici_2004} in the Arp~227 group. It has a projected spatial separation of $5.4\arcmin$ ($\unsim47$~kpc) from its nearest neighbour (the spiral galaxy NGC~470), with both galaxies having identical systemic velocities. While it is clear that the gas-poor NGC~474 is interacting with its gas-rich neighbour \citep{Schiminovich_1997, Cullen_2006, Rampazzo_2006, Mancillas_2019}, the debate about the origin of the observed shell system and its evolutionary link with the ongoing interaction is still open \citep{Schombert_1987, Forbes_1999}. Furthermore, assuming the observed shells have a merger origin, then the nature of the merger event (major/minor) as well as the nature of the disrupted satellite(s) are both unclear.

\begin{figure*}
    \includegraphics[width=0.98\textwidth, scale=0.75]{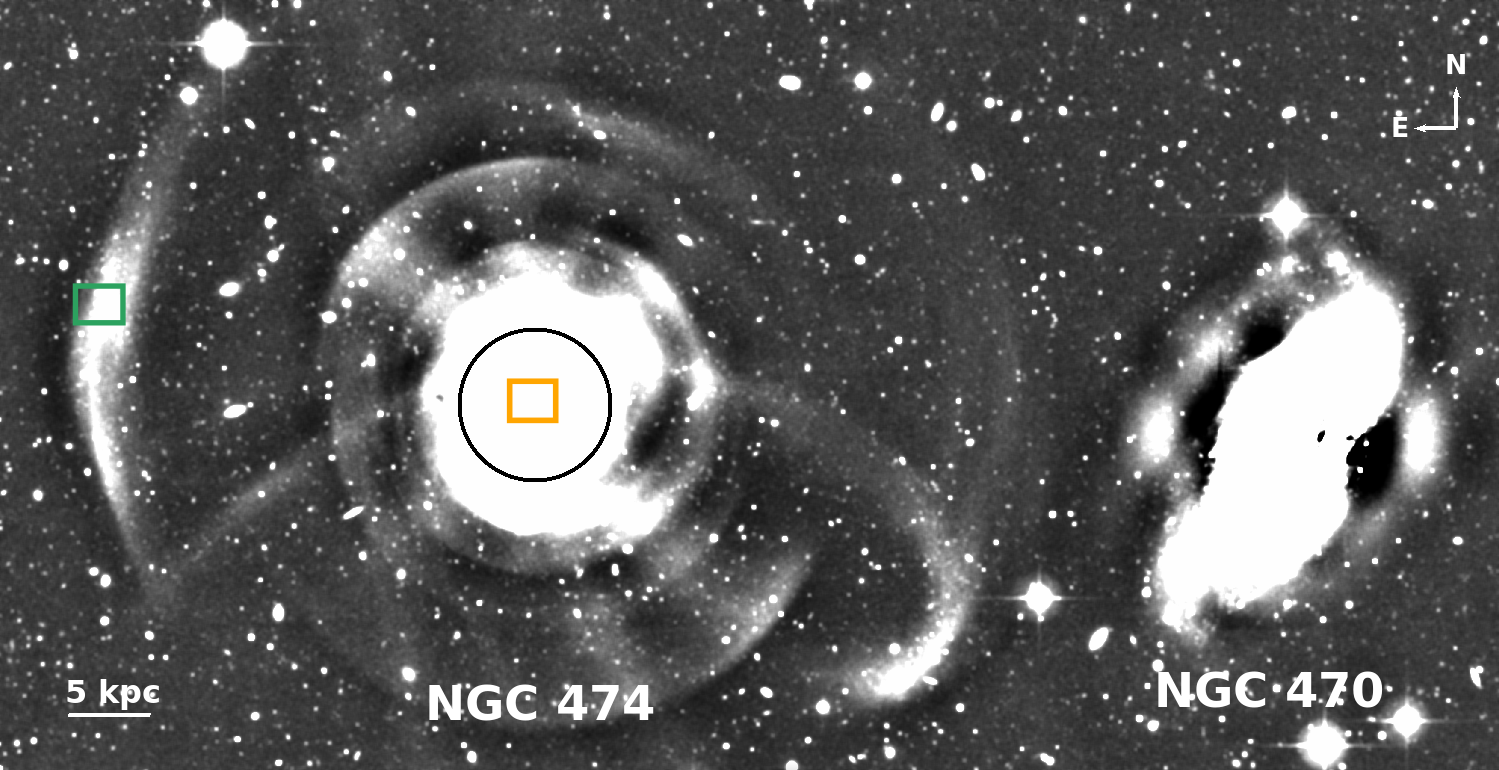}\hspace{0.01\textwidth}\\
	\caption{\label{fig:gal_img} Combined image of NGC~474 from archival CFHT/MegaCam $g, r$ and $i$ images. The image shows the spectacular system of stellar shells and plumes in NGC~474 and covers $11\arcmin \times 5.5\arcmin$ which corresponds to $94 \times 47$~kpc at the distance of NGC~474. The spiral galaxy NGC~470 is at a projected distance of $\unsim47$~kpc away from NGC~474. The orange and green rectangles, each equivalent to $2.4 \times 2.9$~kpc, show the central and shell regions we have observed with Keck/KCWI, respectively. For reference, the black circle is the $1$~effective radius isophote and the white horizontal line has a length of $5$~kpc. The observed shell region is at a physical distance of $\unsim30$~kpc away from the centre of the galaxy.}
\end{figure*}

The central region of NGC~474 has been well studied spectroscopically: with long-slit spectroscopy \citep[e.g.][]{Hau_1996, Caldwell_2003, Denicolo_2005} and with IFU (integral field unit) spectroscopy as part of the SAURON project \citep{Bacon_2001} and the ATLAS$^{\rm 3D}$ survey \citep{Cappellari_2011}. These IFU studies reported an average luminosity-weighted 
age and metallicity of $7.4$~Gyr and $-0.1$~dex within $1$ effective radius ($\re$), respectively, and an $\alpha$--element enhancement, [$\alpha$/Fe], of $0.17$~dex in \citet{Kuntschner_2010} and \citet{McDermid_2015}. NGC~474 is one of the very few fast-rotating galaxies from the ATLAS$^{\rm 3D}$ survey with significant misalignment between its photometric and kinematic axis \citep{Emsellem_2007, Krajnovic_2011}. A similar result was presented earlier from the long-slit spectroscopic study of \citet{Hau_1996}. More generally, spectroscopic studies dedicated to shell galaxies have hitherto focused exclusively on the central galaxy regions \citep[e.g.][]{Longhetti_1998, Hau_1999, Carlsten_2017}, while the stellar shells (usually located in the galaxy outskirts) have not been probed due to their low surface brightness. 

In this paper, we show that direct starlight spectroscopy of stellar shells and streams is now feasible with more sensitive integral field spectrographs such as the new Keck Cosmic Web Imager (KCWI; \citealt{Morrissey_2018}) at Keck. We extend the exploration of the central regions in shell galaxies and build on previous work where discrete tracers such as globular clusters were used to probe shells and streams \citep{Romanowsky_2012, Foster_2014, Lim_2017, Alabi_2020}. We present results from our KCWI IFU spectroscopic observation of NGC~474, covering both the galaxy centre and a region of the outer shell with mean $V$--band surface brightness $22.0$ and $25.4$~\SBunit, respectively. Throughout this work, we adopt a distance of $29.5$~Mpc to NGC~474 \citep{Tully_2013},  $\re=33\arcsec$ or $4.7$~kpc \citep{Cappellari_2011}, and a logarithmic stellar mass of log$(M_*/\Msun)\unsim10.6$ from $3.6$~\micron \ \textit{Spitzer} imaging \citep{Sheth_2010}.

\section{Observations and Data}
We obtained spectroscopic data from the central and shell regions of NGC~474 with the KCWI during the nights of 2017 September 19 and October 18. We used the BM grating centred at $\rm {\lambda}_c = 4800$~\AA \ and the medium image slicer to obtain spectral data with nominal resolution $R\unsim4000$. This configuration spans the spectral region $4360-5230$~\AA~and has a $16.5\arcsec \times 20.4\arcsec$ field-of-view. We therefore obtained data with spectral resolution $\rm FWHM = 1.2$~\AA \ or $\sigma_{\rm instr}\unsim32$~\kms \ at $\rm {\lambda}_c$. We observed the galaxy centre for $600$~s and a region of the prominent eastern shell for a total of $4800$~s, with both regions shown in Figure~\ref{fig:gal_img}. Our observed shell region was chosen to have maximum brightness along the shell but free of compact, point sources such as globular clusters or foreground stars. We also observed a dedicated blank field for accurate sky subtraction.

\begin{figure*}
	\includegraphics[width=0.46\textwidth]{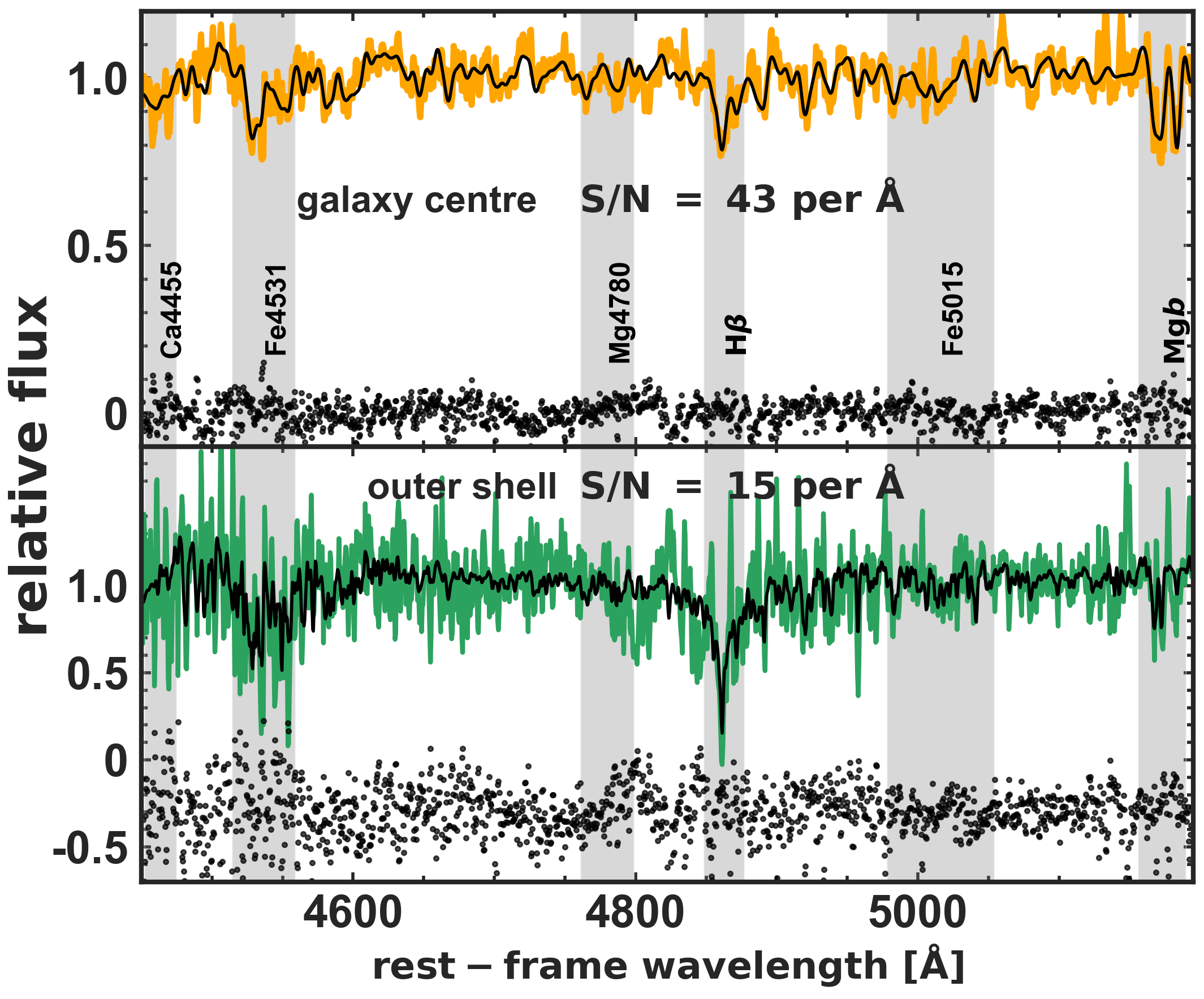}
    \includegraphics[width=0.52\textwidth]{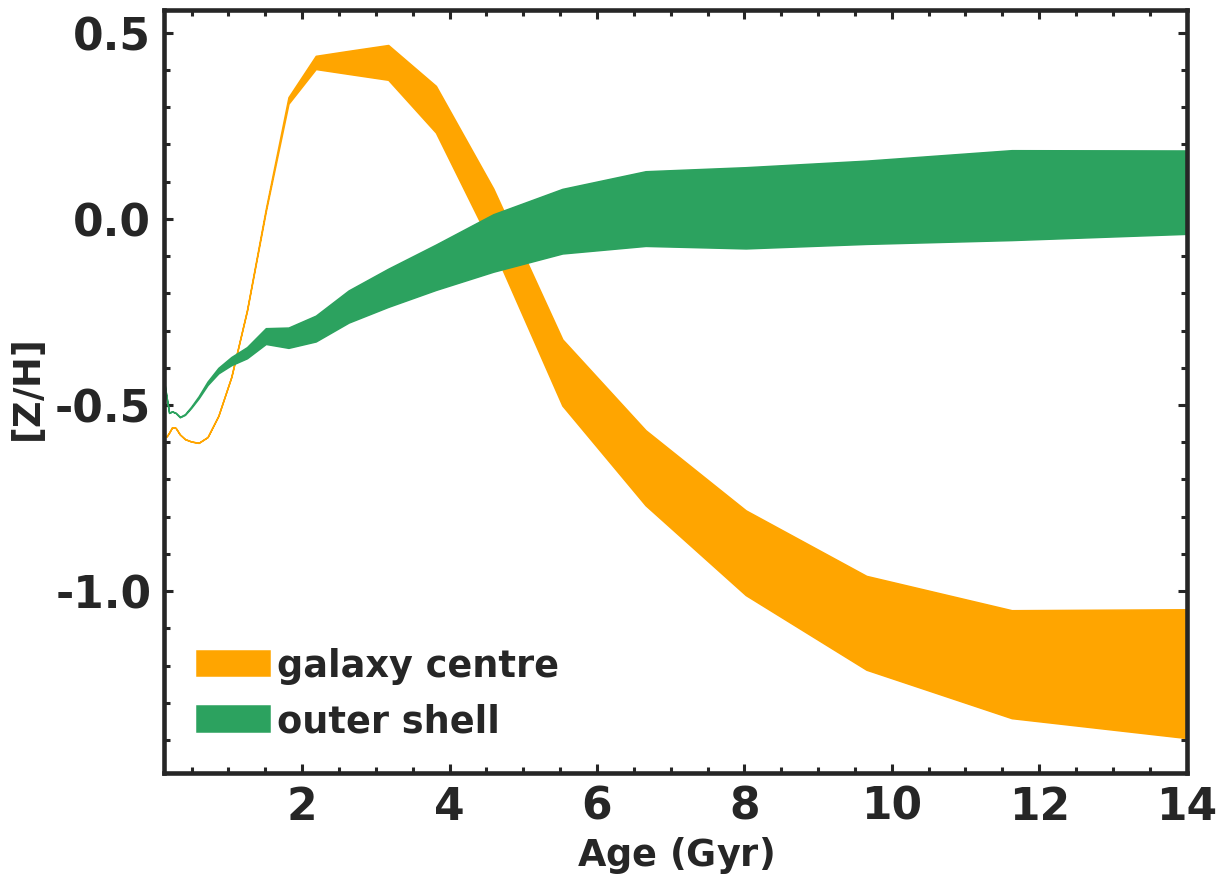}\\
	\caption{\label{fig:gal_spec} Left: Sky subtracted, median normalized Keck/KCWI spectra from the galaxy central region (\textit{upper panel}) and the outer shell region (\textit{lower panel}) of NGC~474, with best-fit {\tt STECKMAP} models shown as black spectra in both panels. Residuals from the best-fit are shown in the lower part of each panel (shifted by $-0.3$ in the lower panel for display purpose only), and the prominent absorption features within the rest-frame spectral range have been identified and labelled. Note the relative strength of the H$\beta$ feature in the \textit{lower panel} compared to  the \textit{upper panel}, hinting at the presence of a substantial, younger stellar population in the outer shell region. Right: Age--metallicity relations from the galaxy centre and the outer shell region. The width of each band corresponds to the stellar mass fractions at different ages. The bands highlight the different chemical evolutionary histories of the galaxy centre and the outer shell region.}
\end{figure*}

Standard data reduction was performed with the publicly available KCWI pipeline -- KDERP\footnote{https://github.com/Keck-DataReductionPipelines/KcwiDRP} -- although we used a custom-made PCA code to perform sky subtraction (see \citealt{Gannon_2020} for details). This is because extra attention has to be paid to sky subtraction especially in the case of spectral data from low surface brightness features such as the stellar shells which are typically $\ge 100 \times$ fainter than the night sky. We collapsed by median-stacking all the useable spaxels within the combined, sky-subtracted fields into single 1D spectra, and obtained final galaxy centre and shell spectra with signal-to-noise (S/N) ratios of $\unsim43$ and $\unsim15$ per \AA, respectively. We show these spectra in Figure~\ref{fig:gal_spec}.

\section{Central and shell stellar kinematics}
We measured the line-of-sight velocities and stellar velocity dispersions at the galaxy centre and from the outer shell region with the upgraded version of the {\tt pPXF} \citep{Cappellari_2004, Cappellari_2017} code. For our kinematics analysis, we employed the empirical ELODIE stellar library \citep{Prugniel_2001} which has a spectral resolution $\rm FWHM = 0.51$~\AA \ (or $\sigma\unsim13$~\kms) \ at $\rm {\lambda}_c$. This allowed us to probe down to (and possibly below) intrinsic velocity dispersions comparable to the instrumental resolution, which may well be the case in the outer shell regions. 
We obtained a radial velocity of $2324\pm6$~\kms \ and a velocity dispersion of $134\pm6$~\kms \ for the central pointing, consistent with measurements in the literature. Note that this velocity dispersion, measured within an $\re/3$ effective aperture, is aperture corrected following the prescription of \citet{Jorgensen_1995}. For the outer shell region, we measured a radial velocity of $2325\pm8$~\kms \ and a velocity dispersion of $18\pm9$~\kms.

\begin{figure*}
	\includegraphics[width=0.44\textwidth]{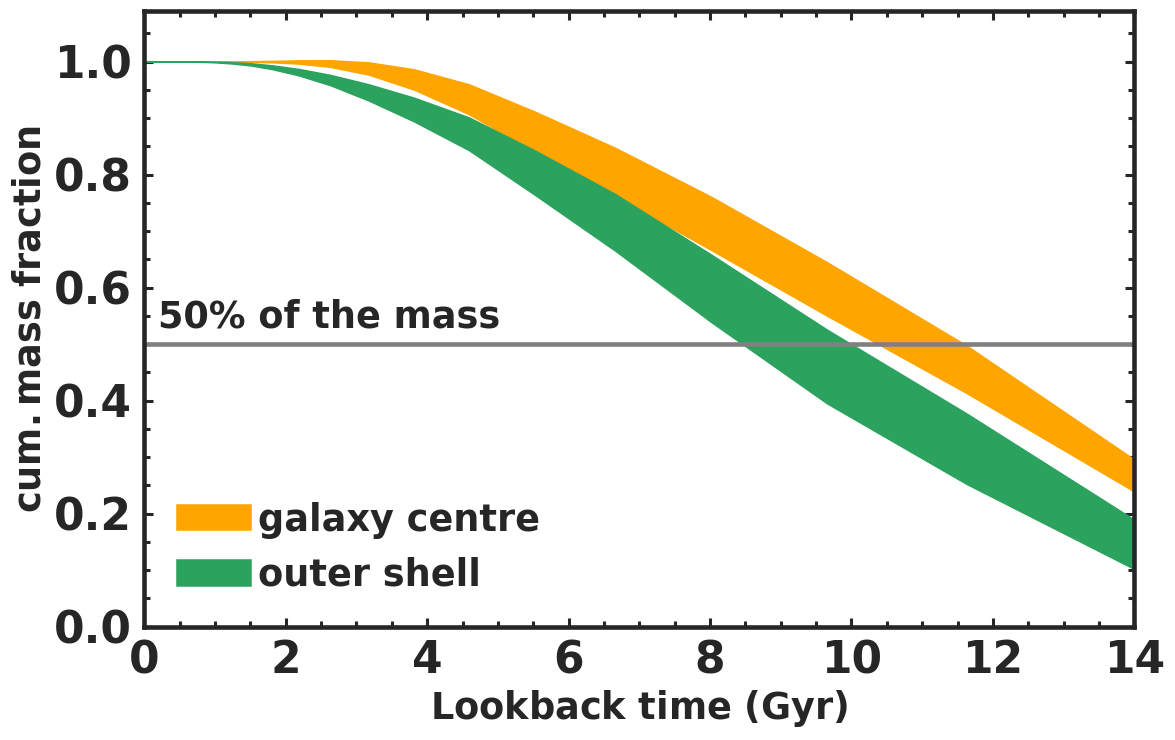}\hspace{0.01\textwidth}
    \includegraphics[width=0.46\textwidth]{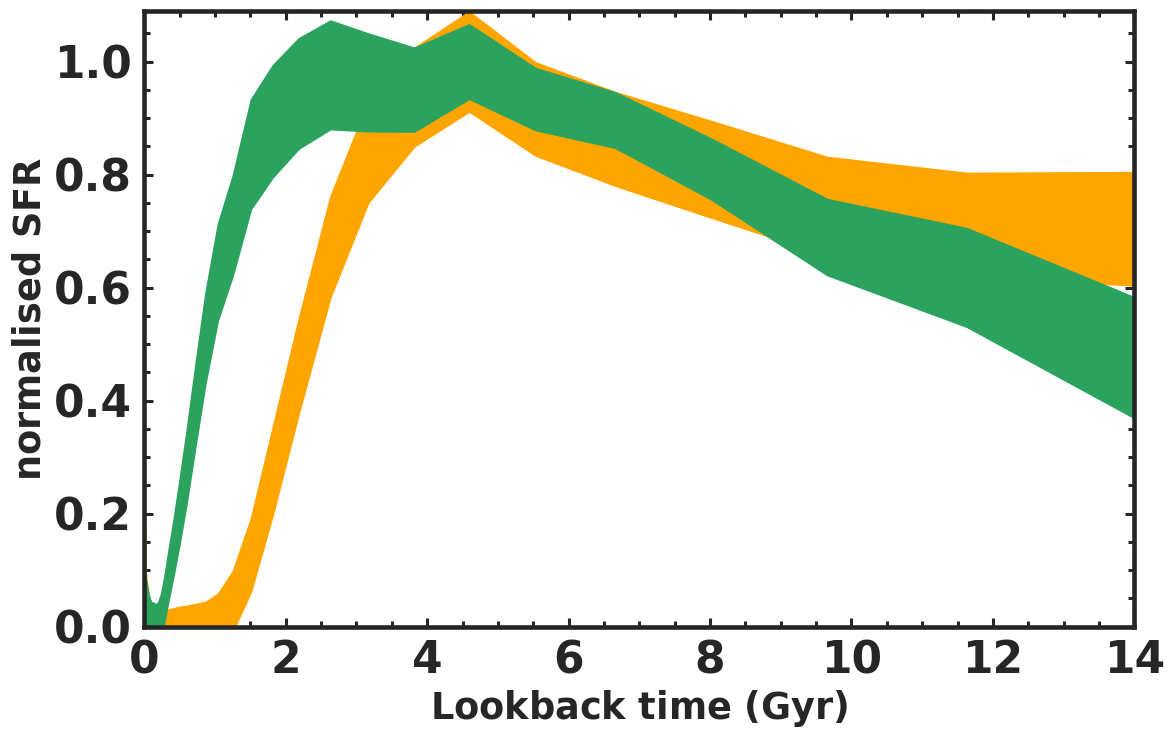}\hspace{0.01\textwidth}\\
	\caption{\label{fig:mass_frxn} Star formation history from the galaxy centre (orange color) and the outer shell region (green color). Both regions of NGC~474 experienced extended star formation but peak star formation at different epochs. The \textit{left panel} shows cumulative mass fraction as a function of lookback time. The orange and green bands show the standard errors around the mean values obtained from Monte Carlo simulations. The horizontal line shows how long it takes to build up 50~per cent of the total stellar mass. The \textit{right panel} shows the normalised star formation rate. The outer shell region experienced a more recent peak of star formation rate (in the last $\unsim2$~Gyr) than the central region of the galaxy.}
\end{figure*}

\section{Stellar Age, Metallicity and Star formation history of NGC~474}
\label{SFH}
As shown in Figure~\ref{fig:gal_spec}, the spectra from both the galaxy central and shell regions include age- and metallicity-sensitive absorption features such as H$\beta$, Fe5015 and Mg\textit{b} (although the latter is on the edge of the spectra). The difference between the outer shell and the central part of the galaxy is already noticeable from their spectra. The outer shell spectrum shows a factor of $\unsim3$ enhancement in the equivalent width of the H$\beta$ feature compared to the galaxy centre spectrum. As H$\beta$ is a strong age indicator, this already suggests the presence of a substantial population of younger stars in the shell region. 

We perform full spectral fitting with the non-parametric regularized {\tt STECKMAP} software \citep{Ocvirk_2006a, Ocvirk_2006b} in order to obtain the mean age and metallicity estimates and the star formation histories from both regions. For this exercise, we use the medium resolution ($R=10,000$) P\'EGASE-HR library of stellar population spectra \citep{LeBorgne_2004, Prugniel_2001} computed with a Salpeter initial mass function at scaled-solar abundances. We model our spectra as a linear combination of these template spectra with ages and metallicities spanning $0.1$ to $14$~Gyr and $-2.0$ to $0.2$~dex, respectively. The best-fitting models and residuals are shown in Figure~\ref{fig:gal_spec}. We adopt regularization parameters $\mu_{x}=100$ and $\mu_{Z}=10$ to smooth the stellar age--metallicity relation shown in Figure~\ref{fig:gal_spec}, and note that the mean stellar population parameters are robust against changes in these values. As suggested by several authors, \citep[e.g.][]{Sanchez_2011, Ruiz_2015}, we fit the stellar population properties separately from the stellar kinematics obtained in the previous section due to possible degeneracies between stellar velocity dispersion and metallicity. We summarize our results in Table~\ref{tab:tab_gal} and provide the mean mass- and luminosity-weighted stellar ages and total metallicities ([$Z$/H]).

As suggested earlier from the strong H$\beta$ feature, the luminosity-weighted age of the outer shell region is $4.0\pm0.5$~Gyr -- reflecting the increased contribution from young stars to the outer shell spectrum -- while the galaxy centre has an older mean luminosity-weighted age of $7.2\pm0.4$~Gyr. In contrast, the mass-weighted stellar ages from both galaxy regions are similar, differing by only $\unsim1$~Gyr. The mean luminosity-weighted ages give a sense of when the last star formation episode occurred whereas the mean mass-weighted age estimates are directly linked to the epoch when the bulk of the stellar mass was formed.

The galaxy centre has a mean mass-weighted [$Z$/H] value of $-0.24\pm0.07$~dex, while the shell region is slightly more metal-rich (by $0.21\pm0.11$~dex, almost at the $2\sigma$ level) with a mean mass-weighted [Z/H] value of $-0.03\pm0.09$~dex. This relatively metal-rich nature of the outer shell region agrees with the metallicity signatures reported in the shell galaxy sample from the Illustris cosmological simulation \citep{Roxana_2017}. We note that the mean luminosity-weighted [$Z$/H] from the shell region is similar to the  measurement from the galaxy centre. This is because luminosity-weighted [$Z$/H] reflects the chemical composition of the underlying old stellar population \citep{Serra_2007}. The central ages and metallicities compare well with previous results from the literature\  \citep{Kuntschner_2010, McDermid_2015} and also agree with the [$Z$/H]--$\sigma$ scaling relation from \citet{Spolaor_2010}.  
 
Figure~\ref{fig:mass_frxn} shows the star formation histories from the galaxy centre and the outer shell region. As expected from fast-rotating, low-mass galaxies in low-density environment \citep{Kauffmann_2004, McDermid_2015, Bernadi_2019}, we obtain evidence of extended star formation from both regions of NGC~474. The galaxy centre formed half of its stellar mass within the first $\unsim3$~Gyr while it took a more extended period of $\unsim5$~Gyr for the outer shell to build up half its stellar mass. Although both regions have extended star formation histories, Figure~\ref{fig:mass_frxn} shows that the shell region experienced a more recent star formation peak within the last $\unsim2$~Gyr. Since our stellar population modelling was done assuming solar abundances, and the lack of a red continuum around the Mg\textit{b} feature makes a direct line-index evaluation of [$\alpha$/Fe] impossible, we use eqn.($3$) from \citet{McDermid_2015} and the time taken to build up half-mass to estimate [$\alpha$/Fe] of $\unsim0.19$~dex and $\unsim0.15$~dex in the galaxy centre and shell region, respectively. This [$\alpha$/Fe] estimate for the galaxy centre is consistent with the results from \citet{Kuntschner_2010} and \citet{McDermid_2015}.

\begin{table}
\centering
\captionsetup{justification=centering}
\begin{tabular}{@{}l c c}
\hline
Parameter & centre & Shell \\
\hline
\hline
$\vlos$~[\kms] & $2324\pm6$ & $2325\pm8$\\
$\sigma$~[\kms]& $134\pm6$ & $18\pm9$\\
Age $_{\rm mass}$~[Gyr]& $8.3\pm0.3$    & $7.1\pm0.5$ \\
$\rm [Z/H]$~$_{\rm mass}$~[dex]& $-0.24\pm0.07$ &  $-0.03\pm0.09$ \\
Age $_{\rm lum}$~[Gyr]& $7.2\pm0.4$    & $4.0\pm0.5$ \\
$\rm [Z/H]$~$_{\rm lum}$~[dex]& $-0.14\pm0.08$ &  $-0.16\pm0.06$ \\
\hline
\end{tabular}
\caption{Summary of results from kinematics and stellar population analyses from both central and shell regions. Parameters with mass and lum subscripts are mass- and luminosity-weighted, respectively.} 
\label{tab:tab_gal}
\end{table}

\section{Discussion}
Our kinematics analysis shows that the radial velocity from the shell region is identical to that of the galaxy centre. Analytical models have shown that the sharp stellar density maxima which define shells are only seen at the apocentres 
along the orbital path of the infalling galaxies, where the mean shell velocity should be comparable to that of the host galaxy \citep{Merrifield_1998, Sanderson_2013}. Our result confirms this observationally and can be used in the future to discriminate bona fide shells from other tidal stellar features when spatially resolved 2D velocity maps become available. From these models, it is not obvious what the velocity dispersion in the shell regions should be relative to the galaxy centre. However, one expects stellar shells to be dynamically cold substructures with low velocity dispersions \citep{Sanderson_2012}. The velocity dispersion we have measured in the outer shell region is lower (at the $\unsim2\sigma$ level) compared to what is measured in log$(M_*/\Msun)\unsim10.6$ galaxies at similar projected radii \citep{Napolitano_2009,Foster_2016}.

Our stellar population analysis shows that NGC~474 has an extended star formation history. This is consistent with expectations for intermediate mass galaxies in low-density environments \citep{Kauffmann_2004}. The galaxy centre is dominated (in luminosity and mass) by subsolar metallicity ([$Z$/H]$=-0.24\pm0.07$~dex and $-0.14\pm0.08$~dex, respectively) intermediate-age stars ($7-8$~Gyr). On the other hand, the stellar population properties of the outer shell region are more complicated. Its age (mass-weighted) and metallicity (luminosity-weighted) are similar to the galaxy centre, but it also hosts a substantial population of younger stars (mean luminosity-weighted age of $4.0\pm0.5$~Gyr) with solar metallicities (mean mass-weighted [$Z$/H]=$-0.03\pm0.09$~dex). The outer shell region is slightly more metal-rich compared to the galaxy centre by $0.21\pm0.11$~dex, significant almost at the $2\sigma$ level. These differences are also captured in the stellar age-metallicity relation of each galaxy region (right panel in Figure~\ref{fig:gal_spec}). The stellar population properties of the stars in the outer shell region therefore suggest an ex-situ origin and an evolutionary path different from the galaxy centre. This also agrees with the conclusions from cosmological simulations \citep{Pop_2018, Mancillas_2019, Karademir_2019}. However, this does not completely resolve the issue of the origin of the shells -- are the shells made from stellar materials accreted from the nearby NGC~470 or are they relics of a disrupted satellite galaxy?

NGC~470 is known to be undergoing an intense nuclear starburst \citep{Friedli_1996} and in a study of the Arp~227 group, \citet{Rampazzo_2006} found HI tidal tails extending from NGC~470 toward the east of NGC~474, in the direction of our shell region (see their Fig. 4). Based on the HI flux in the tidal tails they argued that the gas accretion is in its early stages. Using the maximum rotation velocity of the HI gas in NGC~470 ($240$~\kms), we estimate a timescale of $\unsim0.3$~Gyr for accreted materials to be deposited in the shell region (at a projected distance of $77$~kpc). This timescale is inconsistent with the mean luminosity-weighted age we have measured for the stars in the outer shell, and thus rules out the ongoing interaction with NGC~470 as the sole origin of the observed shell system. Therefore, a disrupted satellite seems to be the more plausible scenario for the formation of the shell system in NGC~474. While we have assumed the simplest merger scenario that involves a single disrupted galaxy, we note that there is observational evidence from the different colors of the various shells in NGC~474 that suggests a multiple merger origin, with further support from cosmological simulations \citep{Pop_2018}. Addressing the question of the multiple merger origin of the stellar shells in NGC~474 is however outside the scope of this study.

The stellar population properties of the stellar shell can shed more light on the origin and nature of the disrupted galaxy. Our analysis above shows that the outer stellar shell is young and metal-rich. Using the galaxy mass--metallicity relation from \citet{Gallazzi_2005}, the outer shell's mass-weighted metallicity of [$Z$/H]$=-0.03$~dex suggests that the disrupted galaxy had a stellar mass of log$(M_*/\Msun)\ge9.6$. This estimate agrees well with the conclusion made by \citet{Lim_2017} that the disrupted galaxy is overmassive for a dwarf galaxy but still sub-$L^*$. \citet{Kim_2012}, using \textit{Spitzer} imaging, already reported that the shells in NGC~474 are $\unsim6-10$~percent as luminous as the underlying galaxy light (see their Table~$4$). If we assume that a single merger event produced the observed shell system and use the finding from the $N$-body simulation of \citet{Hernquist_1992} that shells typically contain $\unsim25$~per cent of the total stars in the disrupted galaxy, then the disrupted galaxy most likely had a stellar mass of log$(M_*/\Msun)\unsim10$. These results show that the shells of NGC~474 most likely originated from an intermediate to major merger event with mass ratio of $1{\,:\,}9$ to $1{\,:\,}3$. Bearing in mind that shells are usually visibile for $3-4$~Gyr after they first appear \citep{Thomson_1991, Pop_2018, Mancillas_2019b} and that mergers are likely to induce star formation \citep[e.g.][]{Barnes_2004}, it is possible that the enhancement in the star formation rate from the outer shell region at $\unsim2$~Gyr is directly connected to the merger event that produced it.

\section{Conclusion}
We have obtained the kinematics and stellar population properties from the galaxy centre and an outer shell region of NGC~474 using new spectral data obtained from Keck with KCWI. We argue that the spectacular stellar shells seen around NGC~474 have an \textit{ex-situ} origin in agreement with expectations from cosmological simulations and were most likely formed in a major or intermediate (mass ratios $1{\,:\,}3$ -- $1{\,:\,}9$) merger event with a metal-rich progenitor $\unsim2$~Gyr ago. Since then, NGC~474 has been accreting cold gas from the outskirts of its gas-rich, spiral galaxy neighbour (NGC~470), which is currently undergoing a starburst. This first, direct spectroscopic observation of the low surface brightness, outer shell region of NGC~474 with Keck/KCWI shows that invaluable insights about the nature of nearby stellar shells in nearby galaxies can now be efficiently obtained on $10$~m telescopes with deep spectroscopy. 
 
\section*{Acknowledgements}
We thank the anonymous referee for the thoughtful reading of the manuscript and for the valuable feedback.
We thank I. Martin-Navarro and A Wasserman for help with the observations.
AFM acknowledges financial support from LCF/BQ/LI18/11630007. 
AJR was supported by National Science Foundation grant AST-1616710 and as a 
Research Corporation for Science Advancement Cottrell Scholar. 
DAF thanks the ARC for financial support via DP160101608. 
ABA and JPB gratefully acknowledge support from National
Science foundation grants AST- 1518294 and AST-1616598.

The data presented herein were obtained at the W. M.
Keck Observatory, which is operated as a scientific partnership among the California Institute of Technology,
the University of California, and the National Aeronautics and Space Administration. The Observatory was
made possible by the generous financial support of the
W. M. Keck Foundation. The authors wish to recognise and acknowledge the very significant cultural role
and reverence that the summit of Maunakea has always
had within the indigenous Hawaiian community. We are
most fortunate to have the opportunity to conduct our research from this mountain.

\section*{Data Availability}
The  observational  data  underlying  this  article  are available online and can be accesed on the Keck Observatory Archive (KOA), which is operated by the W. M. Keck Observatory and the NASA Exoplanet Science Institute (NExScI), under contract with the National Aeronautics and Space Administration.

\bibliographystyle{mnras}
\bibliography{NGC474}

\appendix

\bsp
\label{lastpage}
\end{document}